# Does the derivative of the energy density functional provide a proper quantitative formulation of electronegativity?

Tamás Gál

Department of Theoretical Physics, University of Pécs,
Pécs, Hungary

**Abstract:** It is pointed out that the derivative of the energy density functional does not provide a valid local electronegativity measure, in spite of its appealing property of becoming constant for ground-state equilibrium systems.

Density functional theory [1] has been considered as a natural background for defining chemical reactivity indicators since Parr et al. [2] identified the Lagrange multiplier in the central Euler-Lagrange equation of DFT,

$$\frac{\delta E_v[n]}{\delta n(\vec{r})} = \mu , \qquad (1)$$

as the negative of electronegativity, and proposed a very appealing interpretation of minus the derivative of the energy density functional as a generally local electronegativity measure, which equalizes when an electron system reaches its ground-state equilibrium. This would then provide a formal background for the electronegativity equalization principle [3]. However, so far, the question as to whether $-\frac{\delta E_v[n]}{\delta n(\vec{r})}$ can indeed be considered as a measure of electronegativity even when the given system is not in its ground state (e.g., two molecules before interaction) has not been examined. In the following, we will examine this question, concluding a negative answer.

The electronegativity as defined by the negative of the chemical potential

$$\mu = \left(\frac{\partial E}{\partial N}\right)_{v(\vec{r})} , \qquad (2)$$

appearing as the Lagrange multiplier in Eq.(1), characterizes the change of the ground-state energy induced by a change in the number of electrons in a fixed external potential setting. The main feature of this electronegativity/chemical potential concept, thus, is that it describes energy change due to electron number change. Consequently, its local, non-equilibrium generalization, by

$$\mu(\vec{r}) = \frac{\delta F[n]}{\delta n(\vec{r})} + v(\vec{r}) , \qquad (3)$$

should also characterize energy changes as the electron number changes – but locally.

Consider a functional $F_N[n]$ that equals $F[n]$ for $n(\vec{r})$'s of a given $N$, but otherwise is different from it. This implies that the derivative of $F_N[n]$ with respect to $n(\vec{r})$, at $n(\vec{r})$'s of $N$, differs from that of $F[n]$ by some constant (only),

$$\frac{\delta F_N[n_N]}{\delta n(\vec{r})} = \frac{\delta F[n_N]}{\delta n(\vec{r})} + c . \qquad (4)$$

With $F_N[n]$, then, we have a local quantity

$$\mu'(\vec{r}) = \mu(\vec{r}) + c . \qquad (5)$$

$F_N[n]$ cannot contain information about how the energy (its internal component) behaves when the electron number changes, since $F_N[n]$ may even be constant with respect to changes in the density that go out of the $n_N(\vec{r})$ domain. Consequently, $\mu'(\vec{r})$ cannot be a general, local chemical potential, which characterizes $E$ vs $N$ locally. But then $\mu(\vec{r})$ cannot characterize $E$ vs $N$ locally either, as $\mu'(\vec{r})$ and $\mu(\vec{r})$ differ only by a constant. It *will* characterize something local, but that is not the $N$-dependence of $E$; in other words, it is not $\mu = \partial E/\partial N$ that has been "localized" in Eq.(3). Similar argument holds for the local hardness concept of

$$\eta(\vec{r}) = \left(\frac{\partial \mu(\vec{r})[N,v]}{\partial N}\right)_v , \qquad (6)$$

i.e. [4]

$$\eta(\vec{r}) = \int \frac{\delta^2 F}{\delta n(\vec{r})\delta n(\vec{r}')} f(\vec{r}')d\vec{r}' , \qquad (7)$$

which becomes constant for ground-state densities [5]. That is, on Eq.(6), a hardness equalization principle [6] cannot be based.

In conclusion, density functional derivatives do not provide a good basis for non-equilibrium generalizations of ground-state, constant reactivity indices, such as electronegativity and hardness.